# LifeRaft: Data-Driven, Batch Processing for the Exploration of Scientific Databases

Xiaodan Wang
Dept. of Computer Science
Johns Hopkins University
xwang@cs.jhu.edu

Randal Burns
Dept. of Computer Science
Johns Hopkins University
randal@cs.jhu.edu

Tanu Malik
Cyber Center
Purdue University
tmalik@purdue.edu

## ABSTRACT

Workloads that comb through vast amounts of data are gaining importance in the sciences. These workloads consist of "needle in a haystack" queries that are long running and data intensive so that query throughput limits performance. To maximize throughput for data-intensive queries, we put forth LifeRaft: a query processing system that batches queries with overlapping data requirements. Rather than scheduling queries in arrival order, LifeRaft executes queries concurrently against an ordering of the data that maximizes data sharing among queries. This decreases I/O and increases cache utility. However, such batch processing can increase query response time by starving interactive workloads. LifeRaft addresses starvation using techniques inspired by head scheduling in disk drives. Depending upon the workload saturation and queuing times, the system adaptively and incrementally trades-off processing queries in arrival order and data-driven batch processing. Evaluating LifeRaft in the SkyQuery federation of astronomy databases reveals a two-fold improvement in query throughput.

## 1. INTRODUCTION

Gray and Szalay [9] documented the *data avalanche* problem in the sciences in which improvements in physical instruments and better data pipelines lead to an exponential growth in data size. In Astronomy for example, the Panoramic Survey Telescope and Rapid Response System (Pan-STARRS) produces tens of terabytes daily [10]. Exploring the resulting, massive amounts of data is of immense scientific value.

As scientific repositories scale to petabyte datasets, data become too large to be stored on a single machine. Various scientific disciplines have turned to clustered environments in order to facilitate data exploration. In these clusters, data are typically partitioned spatially or temporally across multiple nodes. These include the Graywulf cluster [2] for the Pan-STARRS Astronomy survey [10], the Turbulence cluster [13] for multi-scale simulation, and Beowulf clusters [23] for geospatial databases. Alternatively, the Sciences have built federations of databases to manage data that are accumulated at multiple, geographically distributed data sources. Examples include SkyQuery [16], Genbank [11], and EcoliHub [5].

Both database clusters and federations achieve high degree of parallelism and aggregate throughput by partitioning data across multiple nodes. This allows scientists to make discoveries by exploring complex relationships at a petabyte scale.

Within the clustered or federated computing environment, a new class of scientific workload has emerged that strains I/O resources. Scientists execute data-intensive join queries to find correlations, mine data, and extract features from datasets that are distributed across multiple nodes. This requires sending tuples between nodes and scanning large portions of the data at each node. Our solution is designed for this computing environment. At each node, we co-schedule concurrent queries that access shared data by executing them in batches against the database server. However, our approach is more suitable for datasets that are partitioned spatially and temporally in which queries can be sub-divided into data-defined units of work that correspond to each partition. These sub-queries are then grouped based on the partitions they access and reordered based on changes in the data access pattern.

Data-driven, batch processing is motivated by SkyQuery. SkyQuery typifies a data-intensive federation with more than a dozen multi-terabyte data sources distributed across three continents. It is facing a scalability crisis in terms of the ever expanding data size and number of sites [24]. Astronomers use SkyQuery to conduct *cross-match* queries that compute spatial joins across distributed data sources [19]. Many cross-match queries have long execution times (several hours or an entire day) and navigate the entire sky, performing full database scans. These queries also transfer large amounts of data over the network.

SkyQuery also receives many short-lived queries, which lead to poor performance with current query scheduling techniques. Short-lived queries (minutes or seconds) focus on a small region of the sky and are highly selective. Database schedulers send queries to the query processor in arrival order, allowing a configurable number (generally tens) of concurrent queries to be executing at any time. In SkyQuery, any scheduler that sends queries to the query processor in order will result in the starvation of short-lived queries that queue awaiting the completion of long-running queries.

Query scheduling needs to be re-examined to ensure high performance in data-intensive scientific clusters and federations. To this end, we provide LifeRaft: a data-driven scheduler that relaxes in-order scheduling to achieve large improvements in query throughput. This is accomplished by exploiting contention between queries for shared data. Specifically, each query is pre-processed to identify its data access requirements. We then read from disk large, sequential regions for which there is high contention and process all queries overlapping that data region in one batch. LifeRaft co-schedules queries that access the same data. This eliminates random and redundant disk accesses. We chose the name LifeRaft because the goal in a life raft is to rescue as many of the most important people as possible. Rescue is not done fairly or in order. To complete the analogy, LifeRaft processes as many concurrent queries as possible against the most contentious data.

To realize starvation resistance, LifeRaft dynamically and incrementally shifts from completing queries in arrival order (low response time) to batch processing (high throughput) based on the workload saturation and queuing times. We employ an *aged workload throughput* metric for scheduling decisions that prefers data regions that have high contention and have queries which have been queued for a long time. A bias parameter weighs the relative importance of contention and age: setting the parameter to 0 selects the most contentious data regardless of age and a setting of 1 completes queries in the order in which they were received. We note that even when evaluating queries in order, the system benefits from data sharing among queries. Thus, when the workload saturates the system, throughput remains high even though queuing times go up and age dominates contention. Balancing contention and age with a bias parameter resembles the seek time versus age trade-off in the VSCAN(R) disk drive scheduler [7].

Instrumenting LifeRaft's data-driven scheduling and starvation resistance in SkyQuery doubles query throughput on data-intensive queries. Owing to higher throughput, data-driven scheduling also reduces response times. Our treatment includes a study of response time versus throughput trade-offs that demonstrates the system's ability to adapt to differing workload saturation.

## 2. RELATED WORK

The query batching paradigm was studied for workloads against large datasets on tertiary storage in order to minimize I/O cost [17, 21, 27]. Yu and Dewitt explored this in the Paradise system by reordering queries over data stored on magnetic tape. The reordering achieves sequential I/O by collecting data requirements during a pre-execution phase (without physically performing the I/O), reordering tape requests, and finally executing queries concurrently in one batch. However, queries participating in the join midway must wait until the entire batch finishes. LifeRaft is not limited to sequential data processing. Sarawagi et. al. [21] provide non-sequential processing by partitioning the data into fragments that are physically contiguous on the tertiary device and scheduling concurrent queries on a per fragment basis. Although our work leverages some of the ideas described in these works, we also explore additional metrics for high query throughput: namely, the amount of data contention and query starvation.

Processing out-of-core spatial join queries has received considerable attention [6, 14, 15, 20]. Patel and DeWitt [20] describe an I/O efficient algorithm for joining relations without spatial indexes that uses a filter step to partitions data based on a minimum bounding box followed by a refinement step that applies computational geometry techniques to join pairs of partitions in memory. Fornari et al. [6] improve the algorithm with uniform object partitioning to prevent overflow in memory constrained environments. Luo et al. [15] proposed parallel algorithms for spatial joins. Whereas a filter and refine approach improves the I/O performance of a single spatial join query, it does not ensure high query throughput over multiple queries. In LifeRaft, we combine the filter and refine approach with batch scheduling that allows for concurrent query execution over shared data. Additionally, we operate on point data that rely on space filling curves, rather than R-trees, for indexing and partitioning.

Google's Map-Reduce [4] is an attractive paradigm for parallel computation and cross-match queries in Astronomy could benefit immensely from parallel execution in this paradigm. Yang et al. [26] extend the Map-Reduce paradigm to more efficiently support relational joins by adding a merge phase that processes heterogeneous datasets simultaneously. Moreover, Olston et al. [18] combined the procedural style of Map-Reduce with declarative SQL constructs in a parallel programming paradigm. Recently, Agrawal et al. [1] incorporated batch processing in the Map-Reduce environment that identifies shared files among map tasks. (Jobs that scan the same files are co-scheduled to maximize throughput). Our work complements their results. However, LifeRaft differs in that caching and memory constraints are crucial considerations for query scheduling. We revisit these differences in Section 6.

We also highlight the current approaches used in SkyQuery to achieve high throughput. The CasJobs [19] system avoids starvation of short queries by data-intensive scan queries through a multi-queue job submission system in which queries from each class are assigned to different servers. The throughput of long running queries is further improved by partitioning the data and evaluating the queries in parallel across multiple servers. However, the distinction between long and short queries is decided arbitrarily and the longest short queries interfere with the short queue and the shortest long queries experience starvation. Our work does not use ad hoc mechanisms to distinguish long and short running queries nor does it rely on multiple servers processing queries for the same data; queries of all sizes are supported in a single system.

## 3. DATA-DRIVEN QUERY PROCESSING

A data-driven approach to query processing focuses on the data requirements of each query, instead of the arrival order, in order to coordinate query processing with access to secondary storage. This is accomplished by first partitioning relational data tables into equal-sized (same number of objects) buckets. Each incoming query is pre-processed to determine a list of sub-queries which satisfy the following property: each sub-query operates on a single bucket and can be processed in any order. The result of the original query is obtained by combining the sub-query results. Finally, buckets are read from disk by scheduler one at a time so that queries whose workload (list of sub-queries) overlaps the bucket are processed concurrently, incurring no additional I/O.

As a first step, we apply the data-driven batch processing to cross-match queries in SkyQuery. (We discuss on-going work that generalize this paradigm to arbitrary join queries in Section 6). A cross-match query is a probabilistic spatial join performed over geographically-distributed Astronomy databases, which record observations of celestial objects. Cross-match is probabilistic as imprecision in physical instruments leads to error in the exact location of an object. It is spatial as astronomers specify an area of the sky for exploration. SkyQuery produces a serial, left-deep join plan for each query that joins (against a large fact table) each archive serially in which intermediate join results are shipped from database to database until all archives are cross-matched. (Cross-match and its operation within SkyQuery are extensively discussed in Malik et al. [16]). In this section, we describe batch scheduling of cross-match queries to permit I/O sharing and maximize throughput.

### 3.1 Partitioning Workload Requests

Currently, cross-match queries in SkyQuery are evaluated exclusively using spatial indices, which is not always beneficial because it introduces random I/O. This is confirmed by Gray et al. [8], which showed that for queries covering a large spatial region, the I/O cost of repeated index access is much higher than a large sequential scan after the application of a coarse filter. They propose a scan-based solution that partitions data into homogeneous buckets, which are sufficiently large (tens of megabytes or more) to amor-tize disk seek times. The partitioning preserves the spatial proximity of objects such that join processing can be localized within each bucket. A coarse filter is applied to each query that assigns cross-match objects to buckets that overlap potential join regions. Then, objects are joined via a sequential pass over the corresponding buckets. While each cross-match object may overlap multiple buckets, each bucket can be joined separately. (Duplicate elimination is not necessary since the spatial join is performed on point data). Our work employs a similar scan-based solution but allows for I/O sharing by batch processing multiple queries.

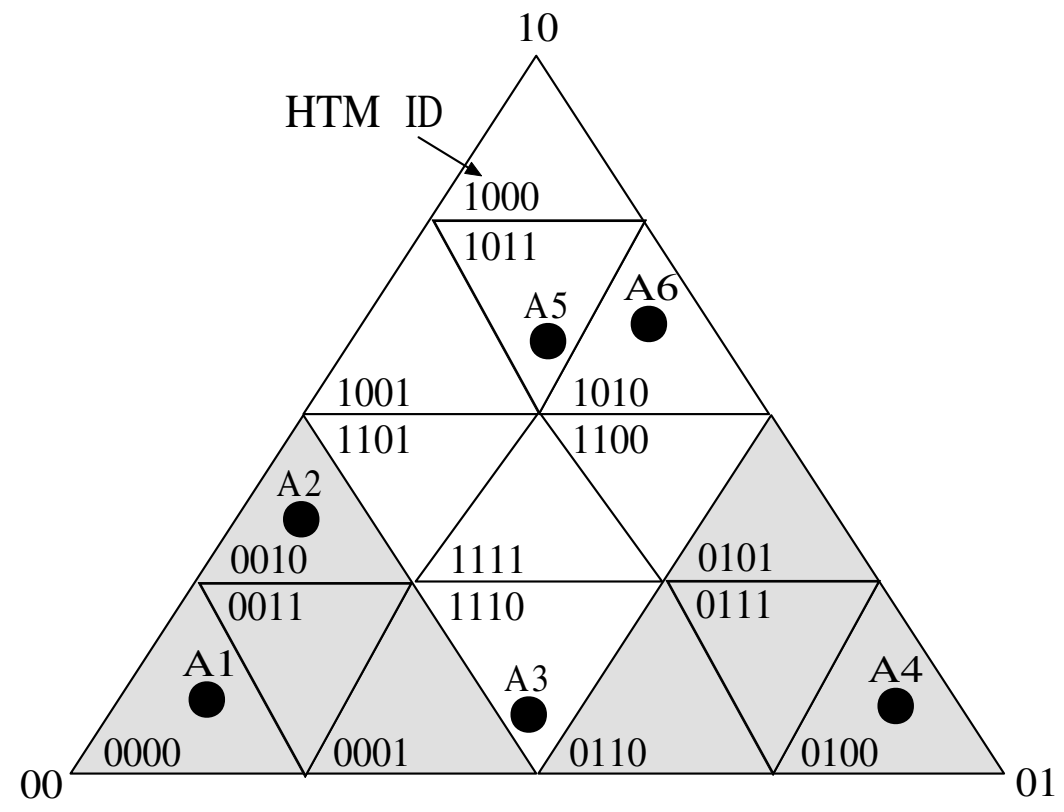

**Figure 1: Defining buckets that contain equal amounts of workload using the hierarchical triangular mesh.**

We employ Hierarchical Triangular Mesh (HTM) to partition data tables [12]. HTM is an efficient method to index objects on a spherical surface through regular quad-tree decomposition of the surface into triangles. It is currently used in SkyQuery to support spatial queries such as computing object intersection and containment. Moreover, HTM is a space filling curve. We employ the latter property to enforce a linear ordering on SkyQuery objects that allows us to partition the data into equal-sized buckets while preserving spatial proximity. Each astronomical observation in SkyQuery is currently assigned a unique 32-bit integer denoting the HTM ID at the fourteenth level in the hierarchy. The numbering preserves spatial locality (objects close in space are also close along the HTM curve) and can be used to derive an object's cartesian coordinates. We partition the sky into disjoint, equal-sized buckets in which each bucket cover a set of triangles that are contiguous in the HTM range (start and end HTM ID values). Equal-sized buckets result in uniform I/O cost for accessing each bucket. Figure 1 illustrates a sample two-level HTM decomposition, which partitions the space into two buckets, each with three objects. Each triangle is labeled with an HTM ID in which two bits is used to denote a triangle's position at each level. The first bucket with HTM IDs `0000` through `0111` contains objects $A_1$, $A_2$, and $A_4$ while the second bucket contains the remaining objects from `1000` through `1111`.

Cross-match objects from multiple queries that map to the same bucket are batched together and joined in a single, sequential pass. Recall that an Astronomy archive receives from each query a list of objects to be cross-matched. Included with each object is its mean cartesian coordinate and a range of HTM ID values, which serve as a bounding box covering all potential regions for cross matching. Let $B_1$, $B_2$, ..., $B_n$ denote the list of buckets and $Q_1$, $Q_2$, ..., $Q_m$ denote the list of cross-match queries at the database. A workload $W_j^i$ represents the set of objects from $Q_i$ that overlap bucket $B_j$ (*i.e.* the object and bucket's HTM ID ranges overlap) and the workload queue for a bucket $B_j$ consists of the union of $W_j^1$, $W_j^2$, ..., and $W_j^m$. Thus, requests from multiple queries are interleaved in the same workload queue and are joined in one pass, thereby minimizing random and redundant I/Os.

We evaluate joins in a data-driven manner by batching together and matching all objects belonging to the workload queue of a single bucket at a time. Objects in both the bucket and its corresponding workload queue are first sorted by their HTM IDs. The join is performed by simultaneously scanning and merging objects in both the bucket and its workload queue. This is similar to the plane sweeping technique used in Partition Based Spatial-Merge Join [20] in which objects are sorted along one axis and then merged. Because cross-match objects from multiple queries are interleaved in the same workload queue, query specific predicates are applied on the output tuples that succeed in the spatial join.

## 3.2 Workload Throughput Metric

Data-driven scheduling maximizes query throughput by biasing toward buckets with more contention in terms of pending workload requests. A longer workload queue means that the cost of reading a bucket from disk can be amortized over more queries. For a bucket $B_i$, we define its *workload throughput* metric as:

$$U_t(i) = \frac{\sum_{j=1}^{m} W_i^j}{T_b * \phi(i) + T_m * \sum_{j=1}^{m} W_i^j} \tag{1}$$

$\sum_{j=1}^{m} W_i^j$ denotes the size of $B_i$'s workload queue. $T_b$ and $T_m$ are constants which estimate the time cost of reading a bucket from disk and the cost of matching a single object in memory respectively. These costs can be derived empirically. Finally, $\phi$ is a function $[1, n] \rightarrow [0, 1]$ in which $\phi(i)$ is 0 if $B_i$ is in memory and 1 otherwise. Buckets are evaluated greedily in order of decreasing workload throughput, which captures the rate in which objects are consumed and cross-matched from its workload queue. The rationale is similar to the selection of paths formed by high throughput links (as measured by the number of packet transmissions per second) in a multi-hop wireless network environment in order to achieve better end-to-end performance [3].

While a greedy policy leads to high query throughput, it may starve requests. Note that scheduling by workload throughput favors buckets in memory that avoid cost $T_b$ of reading from disk. Out-of-core buckets are selected in order of decreasing workload queue length. As a result, the scheduler picks frequently accessed buckets but there is no guarantee that a particular bucket or query receives service. Thus, a balanced approach to scheduling is needed that maximizes throughput while resisting starvation.

## 3.3 Workload Adaptive Scheduling

We describe a starvation resistant scheduler that adapts to workload saturation (query arrival rate) by making trade-offs between response time and query throughput to account for burstiness. A greedy scheduler that maximizes the rate at which objects in the workload queues are consumed benefits overall query throughput, especially for highly saturated workloads. The downside is indefinite starvation of requests, which increases query response time. This trade-off changes for less saturated workloads in which the benefit in query throughput from a greedy scheduler is reduced.

Our solution is inspired by starvation resistant scheduling in modern disk drives [25]. The mechanical nature of disk drives means that locating a piece of data incurs high positioning overhead as measured by seek time and rotational delay. In order to maximize throughput, the scheduler attempts to minimize the movement of the disk head. This is accomplished by accounting for physi-cal characteristics of the media and servicing data blocks that are physically close to the current head position. This favors requests for large, sequentially co-located data blocks and starves requests at other regions on the disk. To counter this, the scheduler accounts for both positioning time to service a request and the wait time of the request. Wait time is used ensure that requests are serviced in arrival order.

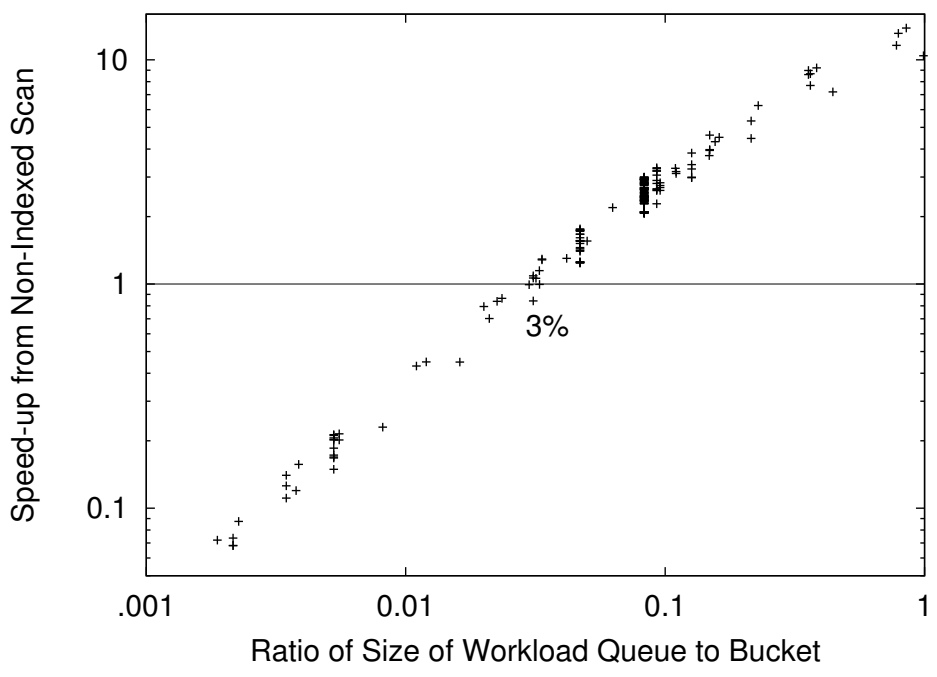

**Figure 2: Speed-up using non-indexed scan vs. spatial index.**

Similar to disk drives, we combine the starvation resistance of processing queries in arrival order with a throughput optimized scheduler by employing the *aged workload throughput* metric, which accounts for query wait times. Starvation can arise for buckets with low workload throughput (small workload queues) or are accessed infrequently. This increases query response time by introducing a last mile bottleneck; namely, a query cannot finish until every object is cross-matched and the remaining buckets with low workload throughput are scheduled. To prevent indefinite starvation, we account for the age $A(i)$, in milliseconds, of the oldest request in each bucket $B_i$. For bucket $B_i$, define its *aged workload throughput* metric as:

$$U_a(i) = U_t(i) * (1 - \alpha) + A(i) * \alpha \qquad (2)$$

$U_t(i)$ denotes the *workload throughput* of a bucket $B_i$ and $\alpha$ is a real value between zero and one, which specifies how strongly biased the scheduler is toward processing requests in arrival order. Parameter $\alpha$ influences the completion order of concurrent queries by specifying an aging criteria on buckets. Even when setting $\alpha = 1$, data sharing occurs among queries. LifeRaft will adaptively tune $\alpha$ based on workload saturation, which we explore in Section 5.

## 3.4 Hybrid Join Strategy

A selective application of indexed join yields performance benefits in the presence of heterogeneous workload distribution. The size of the workload queue for each bucket can vary widely due to differences in query selectivity (*e.g.* differing spatial densities of archives being cross-matched). If indices are available on the join attributes, cross-matching a small workload queue using an indexed join is more efficient because the cost of random I/O accesses is low relative to that of scanning an entire bucket. The performance margin that can be gained or lost depends on the seek time and data transfer rate of the disk.

We employ a hybrid strategy that determines the join plan, either an indexed join or a non-index sequential scan, for each bucket depending on the workload queue size. A pre-determined threshold is used to determine the appropriate join strategy. Figure 2 demonstrates the performance benefits, in SkyQuery, of a non-indexed scan relative to an indexed join as a function of workload queue size. Given buckets that are 40MB in size, we observe up to a twenty fold performance gap depending on the join strategy employed. The break even point occurs when the size of the workload queue is roughly 3% of the size of the bucket.

# 4. ARCHITECTURE

The LifeRaft query scheduler is implemented on top of the database server to support cross-match queries (Figure 3). Incoming queries are first presented to the `Query Pre-Processor`, which takes as input a list of objects to be joined from each query. Based on the spatial attributes of each object, they are assigned to the corresponding workload queues. Workload queues are kept in sorted order based on the *aged workload throughput* metric by the `Workload Manager`. The `Workload Manager` also maintains state information such as a mapping of pending queries to workload queues and the age of the oldest query in each queue.

Once incoming queries are assigned to workload queues, the `LifeRaft` scheduler batches objects from the queue with the highest *aged workload throughput* metric and sends the result to the `Join Evaluator`. The `Join Evaluator` selects the appropriate hybrid join strategy and requests data from the `Bucket Cache`. The `Bucket Cache` either reads an existing bucket from memory or executes a range query to ask for the bucket from the database server. (We use a simple least recently used policy for cache replacement). Finally, the `Join Evaluator` separates objects that succeed in the spatial join by their parent queries, applies query specific predicates, and ships the results to the next site in the cross-match.

Not shown in Figure 3 is the workload adaptive aspect of the *aged workload throughput* metric. We briefly discuss the technique and revisit parameter selection and trade-offs with respect to $\alpha$ in more detail in Section 5.

Parameter selection is based on the query throughput versus response time trade-off curve, which illustrates the effect of adjusting $\alpha$ under a given workload saturation condition. Two such curves are shown in Figure 4 under low (0.1 queries per second) and high (0.5 queries per second) saturation conditions for a sample workload from SkyQuery. (Performance is normalized by the maximum throughput and average response time over all $\alpha$ values). Currently, we determine trade-off curves offline by manually varying workload saturation using a representative workload. The final component is a user specified tolerance threshold, which indicates how much degradation in query throughput is permitted. Figure 4 shows that with an $\alpha$ of 1.0 and 0.25, for low and high saturation respectively, average response time is minimized without sacrificing more than 20% of maximum achievable throughput.

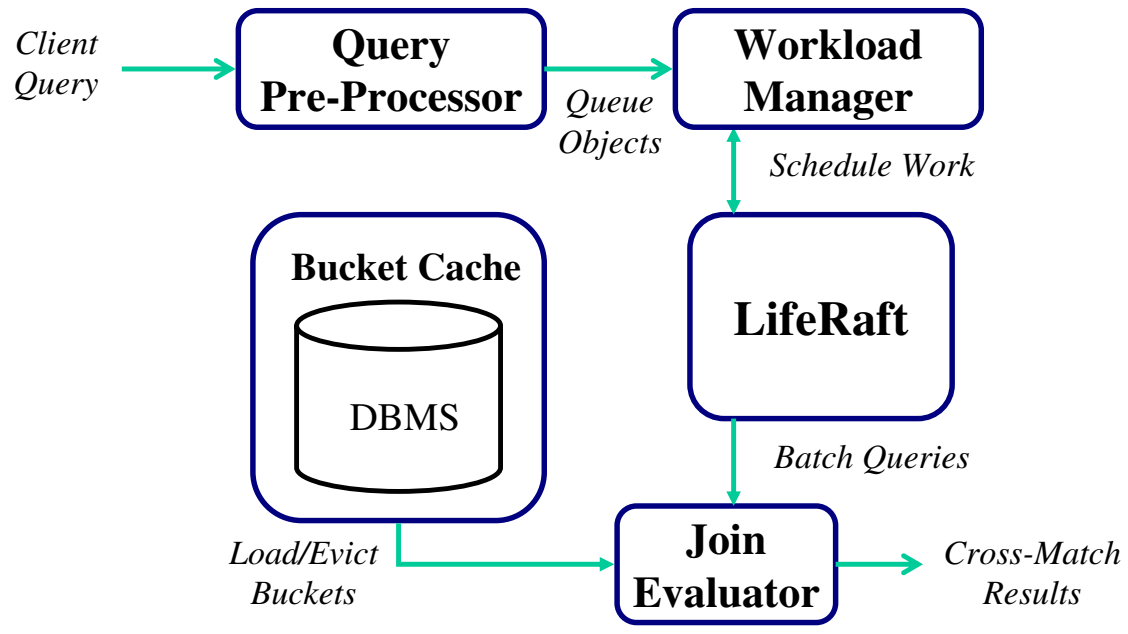

**Figure 3: LifeRaft architecture.**

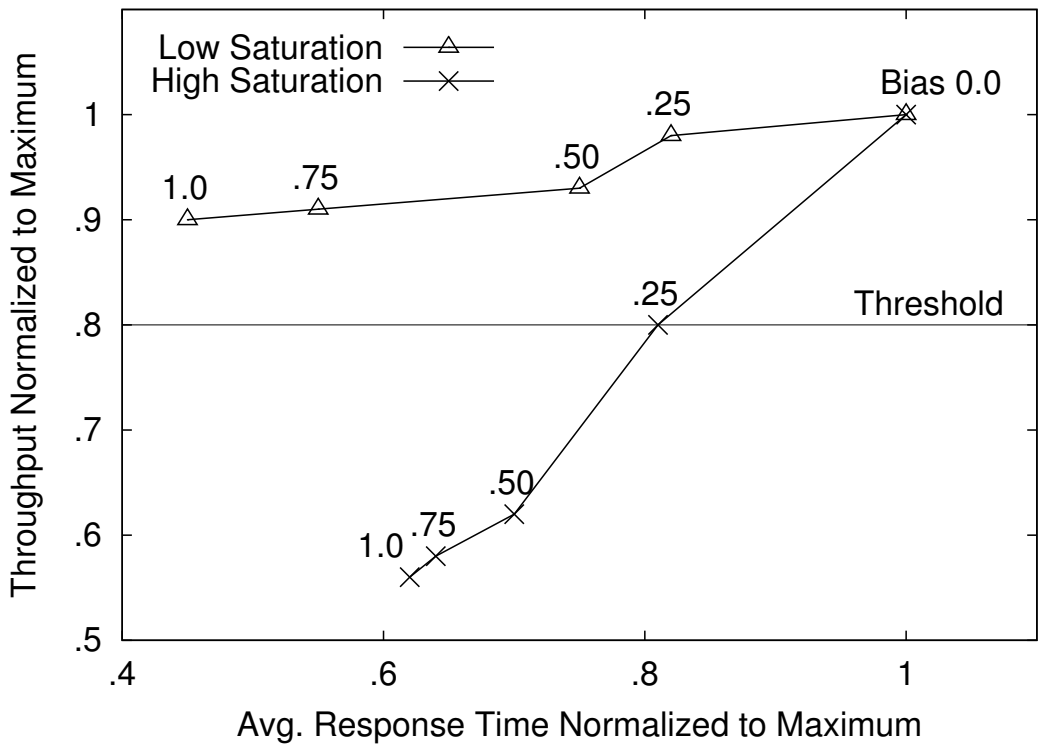

**Figure 4: Performance trade-off curves by saturation.**

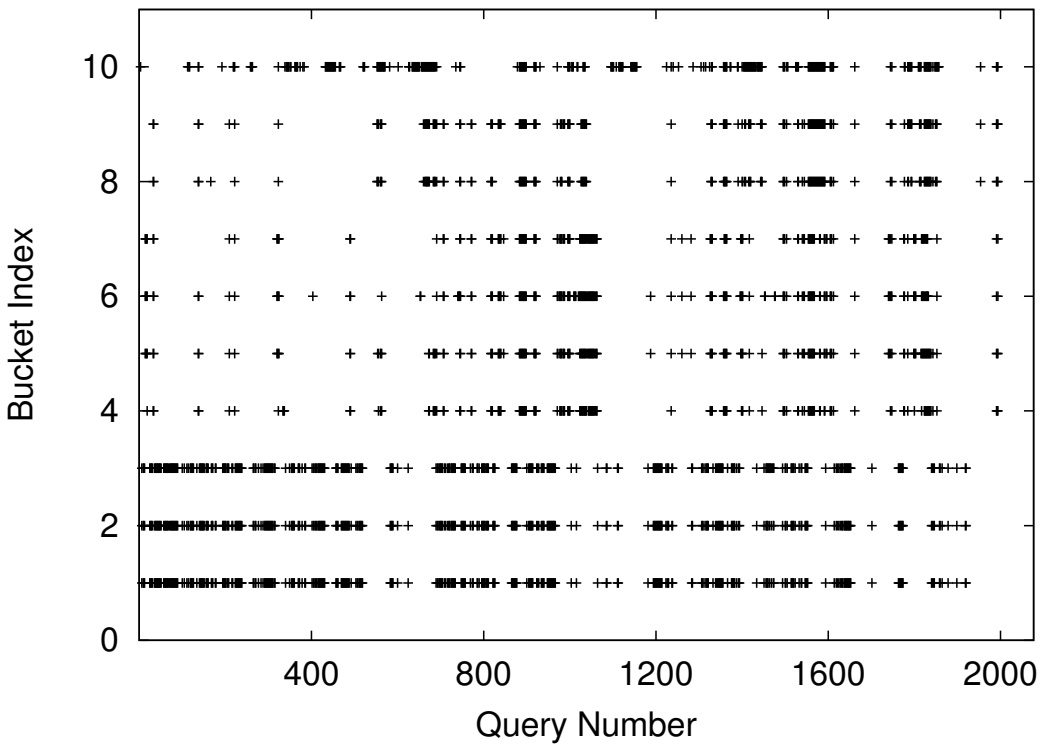

**Figure 5: Top ten buckets by reuse.**

# 5. EXPERIMENTS

We implement LifeRaft, a data-driven query scheduler, in the SkyQuery federation and instrument its performance using a cross-match workload derived from a web log of user queries. We evaluate a range of schedulers by adjusting the age bias for the *aged workload throughput* metric. To measure the benefits of batch processing, we compare with `NoShare`, which evaluates each query independently (no I/O is shared) and in arrival order. We also compare against a round robin scheduler (`RR`), which is a batch processing solution being proposed for SkyQuery. `RR` performs sequential batch processing by servicing buckets in HTM ID order. It is oblivious to both the length of workload queues and age of requests, but is fair in that a request receives the same attention by the scheduler regardless of which bucket it joins with. We do not compare against SkyQuery's existing approach, which evaluates cross-match queries exclusively through spatial indices. (This approach is seven times slower than even `NoShare`).

We evaluate LifeRaft in the Sloan Digital Sky Survey (SDSS) [22], which is a six terabyte archive in SkyQuery. Experiments are performed on a quad-core Windows 2003 server with SQL Server 2005 and 4GB of memory. The data tables are striped across 15 sets of mirrored disks. We partition the primary fact table (on which cross-matching is performed) into 10,000 object buckets, which results in nearly 20,000 buckets that are each 40MB in size. Also, we flush SQL Server's buffer after each bucket is read so that the caching of buckets is managed independently of the database server. (Cache size is fixed at 20 buckets in our experiments). Finally, we empirically derived constants $T_b$ and $T_m$, used in computing workload throughput, as 1.2 seconds and 0.13 milliseconds respectively for the current configuration.

## 5.1 Workload

We employ a two-thousand query trace from SkyQuery consisting of only long running cross-match queries, including ones that were terminated prematurely due to time limits imposed by the system. Focusing only on data intensive, long running queries allow us to conduct a forward looking scalability study while remaining faithful to the science of SkyQuery. Each query may join between two to five database archives, but a vast majority of cross-matches occurs between archives `twomass`, `sdss`, and `usnob`. To measure performance at SDSS, we pre-compute for SDSS intermediate results or list of objects to be cross-matched from each query. This allows us to replay, for each cross-match query, only the work that is performed at SDSS. Further, we assume no workload overflow; memory is sufficiently large to accommodate the cross-match results of each query. Resolving overflow and writing workload queues to disk is left for future work.

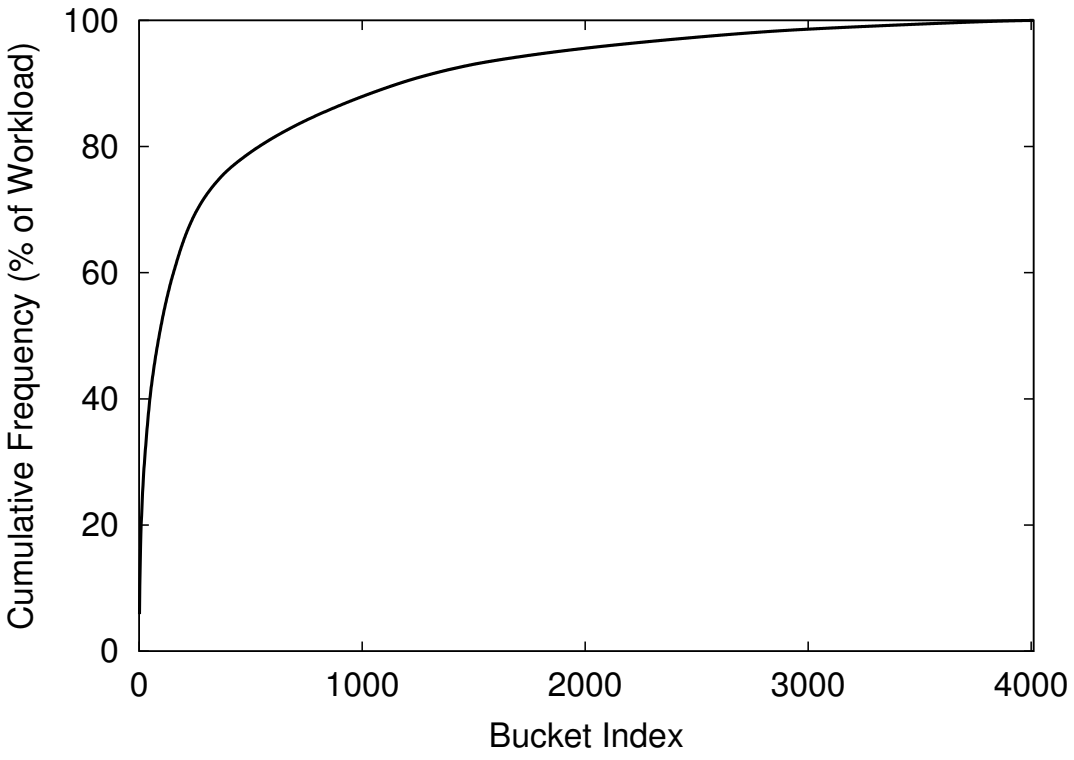

**Figure 6: Cumulative workload by bucket.**

Figures 5 illustrates the suitability of the SkyQuery workload to batch processing and data-driven scheduling. The top ten buckets are reused frequently and accessed by 61% of the queries. Moreover, queries that overlap in data access are close temporally, which benefits caching. Another perspective is shown in Figure 6 in terms of total workload size (measured by the number of cross-match objects) for each bucket. Namely, 2% of the buckets capture 50% of the workload while the remaining buckets that make up the tail are susceptible to starvation by the scheduler.

## 5.2 Results

We compare query throughput and response time of various scheduling algorithms to LifeRaft (varying the value of $\alpha$ for the *aged workload throughput* metric). An age bias $\alpha$ of 0 denotes a greedy approach that schedules buckets in order to maximize query throughput. A bias of 1 schedules buckets in arrival order based on the age of the oldest request. Figure 7(a) shows over two-fold improvement in throughput of the greedy approach over `NoShare`. As $\alpha$ is increased, the performance drop is less pronounced because much of the benefit is derived from batch processing of queries to permit I/O sharing. The performance of `RR` is similar to a LifeRaft scheduler with an $\alpha$ of 1 because neither approach accounts for contention as measured by the sizes of the workload queues.

Figure 7(b) shows the average query response time and variance across various algorithms. `NoShare` exhibits the worst response time, even when compared with the starvation prone, greedy ($\alpha = 0$) LifeRaft scheduler. This is because the I/O overhead of evaluating each query independently leads to higher queuing times. Response time for the greedy approach is nearly twice that of the purely age-based scheduler. This is because queries cannot finish until every object is cross-matched. Since objects may overlap buckets with small workload queues, a query may linger indefinitely until those buckets are scheduled. Biasing toward age leads to incremental improvements in response time at the expense of query throughput. Moreover, `RR` exhibits relatively high average response times. Recall that `RR` schedules buckets by increasing HTM ID order rather than by age of requests. Thus, queries accessing data close to the most recently scheduled bucket (with respect to the HTM ID sequence) are serviced prior to queries that arrived earlier but are further away in the HTM ID sequence. This translates into high delay for certain queries and, in the worst case, forces queries to wait an entire rotation.

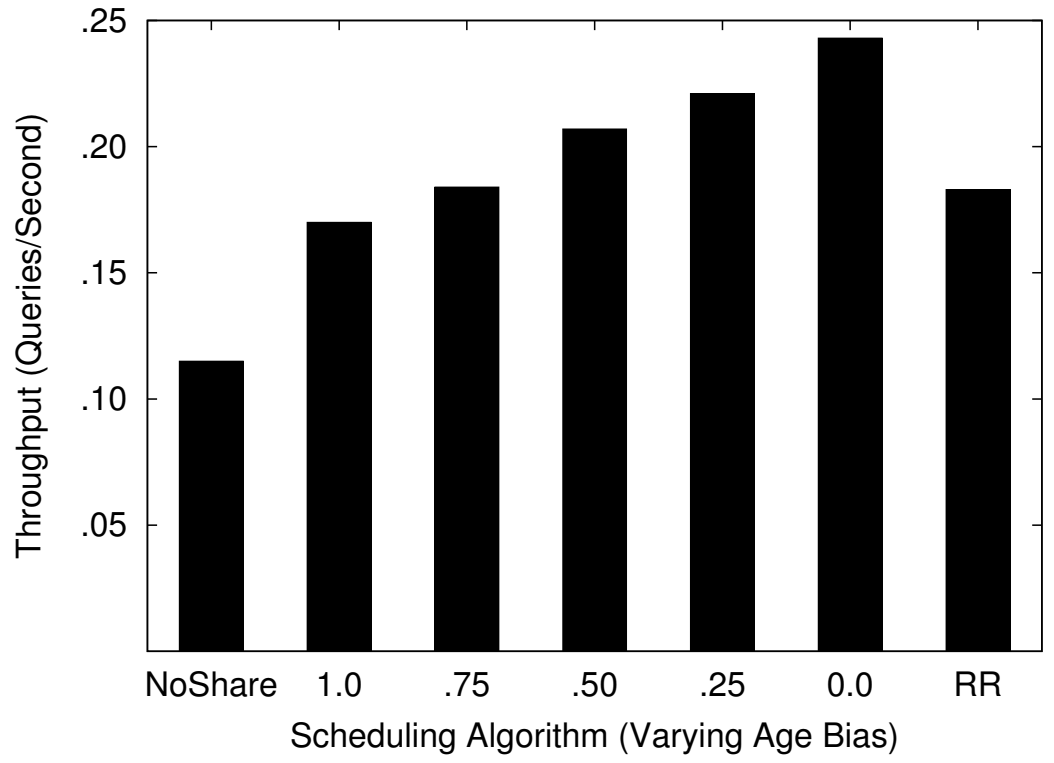

(a) Throughput

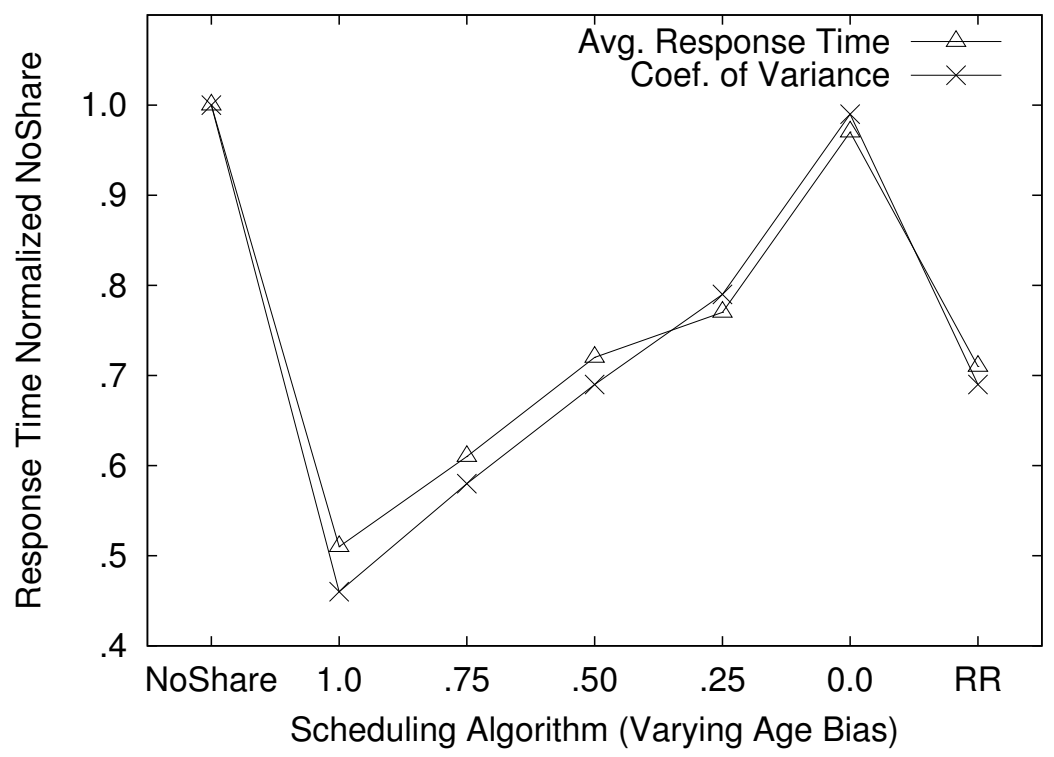

(b) Response Time

**Figure 7: Performance by scheduling algorithm.**

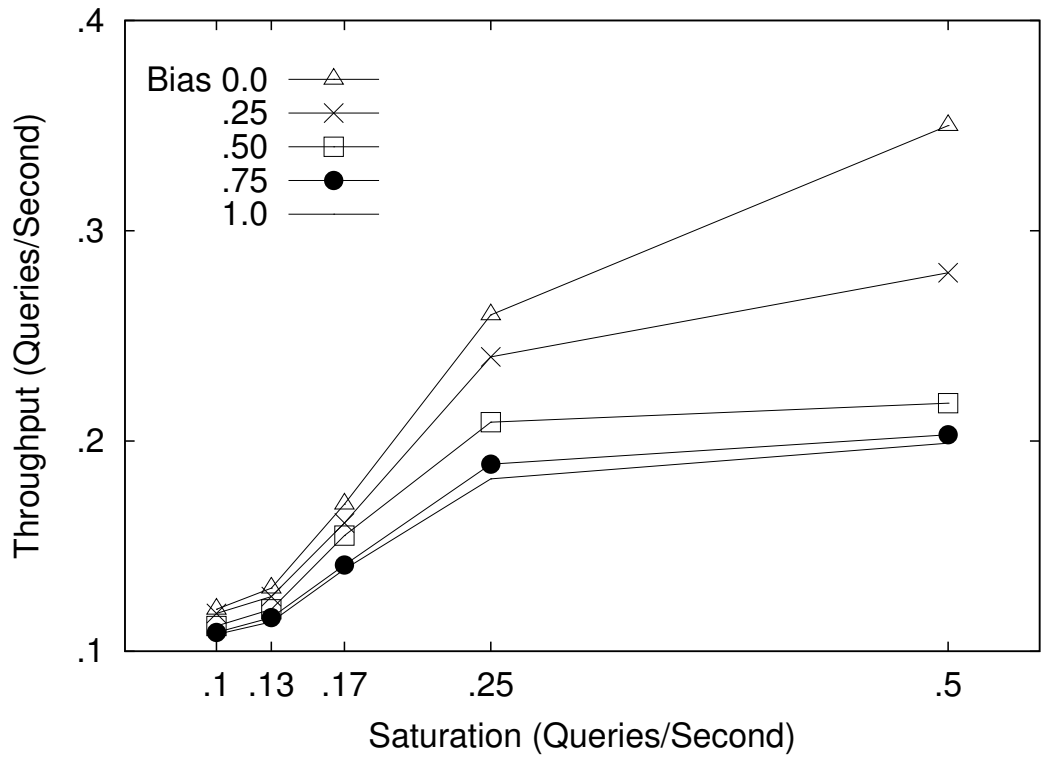

(a) Throughput

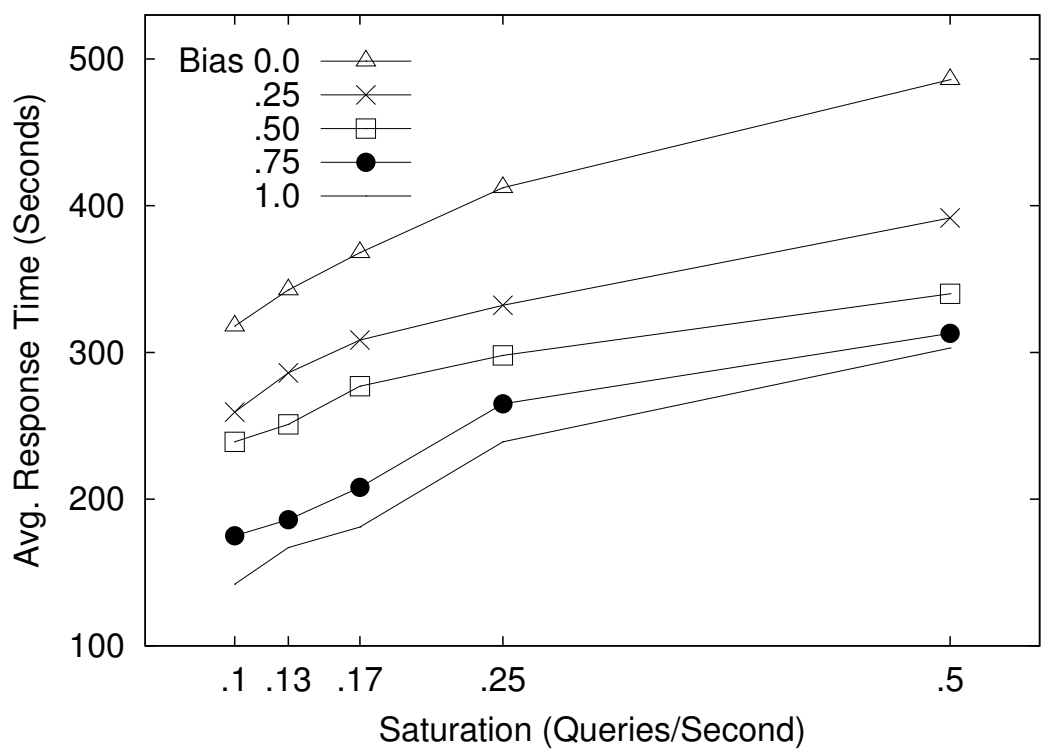

(b) Response Time

**Figure 8: Parameter selection based on workload saturation.**

Figure 8 explores performance trade-offs between query throughput and response time as workload saturation changes. By varying the age bias, the performance gap in throughput (Figure 8(a)) widens considerably as workload saturation changes. This leads to significant improvements in throughput by ignoring arrival order (decreasing the age bias). However, the response time performance gap (Figure 8(b)) remains fairly insensitive to changes in workload saturation. This is due to the hybrid join strategy. Namely, in a contention-based strategy, short-lived queries suffer higher wait times as saturation increases, which is offset by improvements in query throughput performance. The age-based strategy exhibits lower throughput by scheduling short-lived queries that access less contentious data, which may not be present in the cache. To offset the I/O cost of cache misses, these queries are evaluated using spatial indices to avoid scanning the entire bucket. Thus, an age-based scheduler relies more on spatial indices at higher saturations to reduce the overhead of in-order scheduling and maintain low wait times.

In practice, at low saturation conditions, increasing $\alpha$ is attractive. Accepting a small drop in query throughput leads to large improvements in query response time and vice versa. The trade-off curves in Figure 8 help guide the selection of $\alpha$ in real-time. Consider a saturation of $0.5$ queries per second for the current workload and an $\alpha$ of $0$. Increasing $\alpha$ to $0.25$ leads to a 20% drop in throughput and a 20% reduction in response time, which is not an attractive trade-off. However, increasing $\alpha$ to $0.25$ at a saturation of $0.25$ becomes attractive because the same 20% reduction in response time sacrifices only 7% of throughput. In fact, increasing $\alpha$ becomes progressively more attractive with less saturation. (At $0.1$ saturation, a 54% reduction in response time is achieved with only a 7% drop in throughput if $\alpha$ is increased from 0 to 1). Depending on user requirements, such trade-offs can be made more or less aggressive in LifeRaft using the tolerance threshold (Section 4).

# 6. DISCUSSION

We have described the LifeRaft batch scheduler that services data-intensive queries in a data-driven, batch processing paradigm. We have implemented and evaluated LifeRaft for Astronomy's cross-match query. LifeRaft co-schedules the queries that exhibit the highest degree of data sharing with the goal of maximizing query throughput. LifeRaft balances its throughput oriented approach with the need for low query response time for interactive workloads by using an *aged workload throughput* metric. This metric allows the the scheduler to tune batch processing based on workload saturation and user requirements.

Several avenues are open for future work that generalize and improve LifeRaft. Currently, we assume that workload queues fit in memory. In the future, we plan to address workload overflow in which queries will need to be stored to disk and fetched into memory for processing. Queries will be pre-processed to determine the workload queues that join with each bucket. Then, the scheduler will migrate matching pairs of workload queue and bucket into memory for evaluation. We note that in federations, the queries may be quite large, because they include intermediate results from other sites that need to be joined with the current site. We also plan to extend the *aged workload throughput* metric to provide more than just completion order guarantees. Ideally, interactive, short-lived queries should finish quickly and not risk starvation from prior, long-running queries. Thus, we want to provide robust quality-of-service guarantees by depreciating the age bias for longer queries (regardless of the arrival order) to better support both interactive and batch workloads in the same environment.

We plan to provide a theoretical treatment of batch processing by adapting a recent work for shared-scans of files in Map-Reduce systems [1], which formalizes the problem of data sharing among map tasks. They group jobs into batches based on the files accessed so that sequential scans of large files are shared among as many simultaneous jobs as possible. Their solution also includes an aging policy to prevent starvation. However, a direct application to query scheduling is difficult for several reasons. For example, their model is sensitive to job arrival rate, which corresponds to a stationary process. As such, their solution is poorly suited to bursty workloads with no steady state. This is problematic for online systems that serve a continuous stream of queries from tens and thousands of users so that a representative workload is not available. Nonetheless, incorporating arrival rates is an avenue we plan to explore.

LifeRaft uses a different scheduling policy for batch jobs than the system for shared file scans in Map-Reduce systems. Agrawal et al. [1] rely on a least sharable file first policy in which jobs that do not benefit from co-scheduling with future jobs are executed first. This is in contrast to our most contentious data first policy, which we expect to perform better on scientific workloads. One reason is that, unlike Map-Reduce, buffering workload queues (intermediate join results) requires significant amounts of storage. Given that many queries are already I/O bound, accumulating large amounts of intermediate results in memory and writing them to disk, which allows for larger batch sizes later, is undesirable. Our *workload throughput* metric addresses this problem; servicing the most contentious region first, which reduces buffering requirements.

Caching also benefits a contention based scheduler. In the Map-Reduce environment, shared files are too large to fit in memory such that only one file is scanned at a time. This is analogous to a cache size of a single bucket. However, keeping multiple buckets in memory means that future queries referencing contentious data regions benefit from reuse and incur no additional I/O. Not surprisingly, in comparing the most data-sharing ($\alpha = 0$) policy with a purely age-based scheduler ($\alpha = 1$), we found 40% and 7% of requests serviced from the cache respectively. This is because an age-based scheduler may evict contentious data regions to maintain completion order. We plan to confirm the advantages of caching and less buffering empirically in future work.

Finally, integrating batch processing in a distributed query processing environment presents additional challenges. Our solution allows individual sites in a cluster or federation to batch queries independently, but it is not clear whether coordinating schedules across multiple sites is beneficial. The latter requires that every site is aware of the data access requirements of each query *before* queries visit and join these sites. The advantage being that different sites can coordinate query execution order to maximize the batch size over all sites (*e.g.* amortize I/O cost over more queries). In this setting, the least sharable data first policy makes sense; a site will delay processing of a bucket if it anticipates workload that is pending at another site and accesses the same bucket.

LifeRaft has applicability beyond Astronomy to other scientific, data-intensive workloads. The benefits of batch processing for large database federations, such as SkyQuery, are obvious. In SkyQuery, millions of queries are processed each month so that scalability, as measured by query throughput, is a crucial performance metric. The same argument applies to many other applications as they evolve to larger data sets and workloads. Specifically, spatio-temporal databases [13] in which queries can be sub-divided into data-defined units of work that operate on static, non-overlapping data partitions. However, regardless of the application, batch processing must be mindful of diverse user requirements so that batch workloads do not starve interactive, short-lived queries.

## Acknowledgments

The authors wish to thanks Alex Szalay who contributed substantially to both the vision and technical details of this work. We also wish to thank Ani Thakar, Tamás Budavari and the rest of the Sloan Digital Sky Survey team at Johns Hopkins University for their assistance with Astronomy data and workloads. This work was supported in part by NSF awards IIS-0430848, AST-0428325, and ATM-0618986.